\documentclass{aa}  
\usepackage{natbib}
\usepackage{graphicx}
\usepackage{txfonts}

\newcommand{\sax}{{\it {\mbox BeppoSAX}}\xspace}
\newcommand{\ledd}{L_{\rm Edd}\xspace}
\newcommand{\Itot}{I^{\rm tot}\xspace}

\newcommand{\Fplus}{F^{(+)}\xspace}
\newcommand{\Fminus}{F^{(-)}\xspace}

\newcommand{\Iup}{I^{(+)}\xspace}

\newcommand{\Iuplk}{I^{(+)k}_{\rm l}\xspace}
\newcommand{\Iuprk}{I^{(+)k}_{\rm r}\xspace}
\newcommand{\Iuplkminusone}{I^{(+)k-1}_{\rm l}\xspace}
\newcommand{\Iuprkminusone}{I^{(+)k-1}_{\rm r}\xspace}

\newcommand{\Iminus}{I^{(-)}\xspace}
\newcommand{\Iminusl}{I^{(-)}_{\rm l}\xspace}
\newcommand{\Iminuslr}{I^{(-)}_{\rm l,r}\xspace}

\newcommand{\Iminuslk}{I^{(-)k}_{\rm l}\xspace}
\newcommand{\Iminusrk}{I^{(-)k}_{\rm r}\xspace}
\newcommand{\Iminuslkminusone}{I^{(-)k-1}_{\rm l}\xspace}
\newcommand{\Iminusrkminusone}{I^{(-)k-1}_{\rm r}\xspace}

\newcommand{\Ikrl}{I^k_{\rm r,l}\xspace}
\newcommand{\Ikl}{I^k_{\rm l}\xspace}
\newcommand{\Ikr}{I^k_{\rm r}\xspace}
\newcommand{\Il}{I_{\rm l}\xspace}
\newcommand{\Ir}{I_{\rm r}\xspace}

\newcommand{\Itotrlup}{I^{(+){\rm tot}}_{\rm r,l}}
\newcommand{\Itotrldown}{I^{(-){\rm tot}}_{\rm r,l}}

\newcommand{\Itotrup}{I^{(+){\rm tot}}_{\rm r}}
\newcommand{\Itotlup}{I^{(+){\rm tot}}_{\rm l}}

\newcommand{\blfreq}{\nu^{\rm bl}\xspace}
\newcommand{\lx}{L_{\rm X}\xspace}
\newcommand{\kte}{kT_{\rm e}\xspace}
\newcommand{\ftot}{F^{\rm tot}\xspace}

\newcommand{\Rns}{R_{\rm ns}\xspace}
\newcommand{\Rbl}{R_{\rm bl}\xspace}
\newcommand{\Rg}{R_{\rm g}\xspace}
\newcommand{\mui}{\mu_{\rm i}\xspace}

\newcommand{\thetamax}{\theta^{\rm max}\xspace}

\newcommand{\cygx}{\mbox{Cyg X-2}\xspace}
\newcommand{\scox}{\mbox{Sco X-1}\xspace}
\newcommand{\gxnine}{\mbox{GX 9+9}\xspace}
\newcommand{\xtej}{\mbox{XTE~J1701--462}\xspace}
\newcommand{\scw}{Schwarzschild\xspace}

\newcommand{\dsigmaprime}{{\rm d}\Sigma^{\prime}\xspace}
\newcommand{\Stot}{{\bf \mathcal{S}}^{\rm tot}}

\bibpunct{(}{)}{;}{a}{}{,} 

\usepackage[normalem]{ulem}

\begin{document} 

   \title{The polarization of the boundary layer around weakly magnetized neutron stars in X-ray binaries}
   \titlerunning{Polarization of the boundary layer in X-ray binaries}

   \author{R. Farinelli
          \inst{1}
          \and
          A. Waghmare\inst{2}
          \and
          L. Ducci\inst{2,3}
         \and
          A. Santangelo\inst{2}          
          }

   \institute{INAF -- Osservatorio di Astrofisica e Scienza dello Spazio, Via P. Gobetti 101, I-40129 Bologna,  Italy \\
              \email{ruben.farinelli@inaf.it}
         \and
             Institut für Astronomie und Astrophysik, Kepler Center for Astro and Particle Physics, University of Tübingen, Sand 1, 72076 Tübingen, Germany
         \and
            Department of Astronomy, University of Geneva, Chemin d'Ecogia 16, 1290 Versoix, Switzerland\\
                 }

\date{Received December 15, 2023; accepted January 16, 2024}
  
\abstract
   {X-ray binaries hosting a compact object have been among the main targets of the Imaging X-ray Polarimetry Explorer (IXPE)  since its launch, due to their high brightness in the 2-8 keV energy band. The spectropolarimetric analysis performed so far has proved to be of great importance in providing constraints on the accretion geometry of these systems. However, the data statistics is not enough to unambiguously disentangle the contribution of the single components to the net observed polarimetric signal.}
   {In this work, we aim to present a model for computing the polarization degree and polarization angle of the boundary layer around 
   weakly magnetized neutron stars in low-mass X-ray binaries in the soft state. The main motivation is to provide strong theoretical support to data interpretation of observations performed by IXPE or future satellites for X-ray polarimetry.}
   {The results were obtained by modeling the boundary layer as an equatorial belt around the compact object and locally approximating it as a plane-parallel scattering atmosphere, for which the associated
   radiative transfer equation for polarized radiation in the Thomson limit was solved.
   The polarimetric quantities were then transformed from the comoving frame to the observer frame using the numerical methods formerly developed for X-ray pulsars.}
   {For typical values of the  optical depth and electron temperature of the boundary layer of these systems in a soft state, the polarization degree was less then 0.5\%,  while the polarization angle was rotated by $\protect\la 5^{\circ}$ with respect to the neutron star spin axis due to special and general relativistic effects for fast rotation, the amount progressively decreasing for lower spin frequencies. The derived quantities can be used to remove degeneracy when multicomponent spectropolarimetry is performed.}
   {}


\maketitle
  
\section{Introduction}\label{sect_introduction}
Low-mass X-ray binaries hosting a weakly magnetized neutron star (NS-LMXBs) are among the brightest sources of
the X-ray sky and have been studied since the dawning of X-ray astronomy. 
In the Milky Way, about 150 NS-LMXBs are known \citep{avakyan2023}.
They have historically been divided into two main classes -Z and atoll sources- on the basis of the shape they track
in color-color or color-intensity diagrams \citep{hasinger1989}.
The main characteristics of Z sources are a persistently bright luminosity ($\lx \ga \ledd$) and
a relatively stable spectrum that changes slightly while the source moves along the Z track.
The X-ray spectrum is usually modeled in terms of a soft thermal component,
plus an unsaturated Comptonization spectrum of $\sim$ 1 keV black-body-like photons  off a thermal
electron distribution with $\kte \sim$ 3--5 keV in a high optical depth environment and $\tau \sim$ 5--10 depending
on the assumed geometry, slab, or spherical, respectively \citep{disalvo2000, disalvo2001, damico2001, lavagetto2004, farinelli2008, farinelli2009, homan2018}.
Additional features are the presence of an iron fluorescence line, as well as a transient, hard X-ray tail
above 30 keV, which appears correlated to the radio activity and in turn the formation of a jet
\citep{paizis2006}.
On the other side, atoll sources cover a wider range of luminosities 
(from $\la 10^{-3}$ up to $\sim \ledd$), sharing in this fashion a common property with black-hole X-ray binaries
\citep{done2007}.
The changes in luminosity are linked to variations of the spectral state, with the brightest
luminosity achieved when the X-ray spectrum is Z-like (soft state), and dimmer fluxes (hard state) in which the spectrum
is characterized by $\kte \sim 30$ keV and $\tau \sim$ a few \citep{gierlinski2002, 
maccarone2003, falanga2006, cocchi2011}.
There is also a subclass of bright atoll sources (GX 9+9, GX 9+1, GX 3+1, and GX 13+1) that have
never shown spectral transition, with a rather stable X-ray spectrum typical of Z sources \citep{mainardi2010, iaria2020}. 

For a long time, understanding the shape of the accretion process in NS-LMXBs has been a challenge: in particular, it has long been discussed about the hydrodynamical configuration of the inner region responsible for the strong, dominating, thermal Comptonization bump. It has been proposed as either a diffuse accretion disk corona \citep{white1982} or the region
between the inner accretion disk and the NS surface \citep{frank2002}. In the latter case, in the literature it is often
defined as the boundary layer (BL); albeit, in some cases it is also called
spreading layer \citep{inogamov1999, suleimanov2006}.

The origin of the direct thermal emission at lower energies has been a matter of debate as well.
Historically, this has led  to spectral modeling degeneracy, in which the observed X-ray spectrum could be equally well described
by an unsaturated Comptonization component plus either a direct black-body (BB) emission, attributed to the NS surface
\citep[Western Model, ][]{white1988}, or a direct contribution from the accretion disk \citep[Eastern Model, ][]{mitsuda1989}.
Somewhat surprisingly, even missions with unprecedented broad-band coverage such as \sax \citep{boella1997} have
not been able to remove the degeneracy.
A strong boost in favor of the Eastern Model scenario has been obtained from the analysis of a
sample of atoll and Z sources using Fourier frequency-resolved spectroscopy of archival  data of the Rossi X-Ray Timing Explorer \citep{gilfanov2003, Revnivtsev06,Revnivtsev13}.
In these works, the authors have shown that the unsaturated Comptonization region can be reasonably described
in terms of a spreading layer covering the NS surface in a  belt-like fashion, with a power spectrum resembling the Fourier frequency-resolved spectrum at the period of quasi-periodic oscillations. 
\indent Significant expectations for a major leap in knowledge  were put on the Imaging X-ray Polarimetry Explorer  \citep[IXPE,][]{weisskopf2022}, because the energy range of the spacecraft (2--8 keV) covers a substantial fraction of the  
emission bulk in NS-LMXBs, in particular in the region where the soft and hard components overlap.
Indeed, X-ray polarimetry does constitute the effective third axis of an ideal vector in which the other
two components are the spectral and temporal analysis, respectively, and it can be fundamental in removing model
degeneracy.
Joint observations of IXPE with other X-ray facilities covering a wider energy range have shown how spectropolarimetric information can provide unprecedented constraints on the inner region of accreting X-ray binaries \citep[see e.g., ][for the black hole Cyg X--1]{krawczynski2022}.
A sample of NS-LMXBs in soft state have been observed so far by IXPE; however, despite some very intriguing results, the data
statistics have not allowed us to put unambiguously tight constraints on the polarization degree (PD) and  polarization angle (PA) of the different components that contribute to the total observed spectrum.  \newline
 \indent Different values of the PD of the BL component have been observed when performing model-dependent spectropolarimetric analysis. Concerning Z sources, for \cygx \cite{farinelli2023} found a value of $\approx 4\pm 2$\% during a possible normal branch (NB) phase, while when studying  GX~5--1  \cite{fabiani2023}
obtained $5.7 \pm 1.4$\%  in the horizontal branch (HB) 
 and  $4.3 \pm 2.0$\% in the state time-integrated over the NB and flaring branch (FB). However, in the second case
  the source showed a peculiar behavior with the presence of a direct BB component plus a Comptonization feature.
For \scox, the PD obtained by \cite{lamonaca2023} was $1.3 \pm 0.4$\% when the source was in the soft apex joining the NB with the FB.
Studying the outburst of the peculiar source \xtej, \cite{cocchi2023} modeled the IXPE-alone X-ray spectrum
with a simple BB + multicolor BB disk model and derived a PD of $\approx 6 \pm 0.5$\% 
for the hard BB component when the source was in the HB, the value dropping to $\approx 1.2 \pm 0.7$\% 
in the NB.  \newline
\indent Atoll sources in a soft state have been investigated as well. In the case of the bright source \gxnine \citep{ursini2023}, the presence of a reflection component, necessary from the spectral point of view, added further degeneracy
in the spectropolarimetric analysis. 
Among different combinations, the authors fixed the PD of the reflection to 10\%, obtaining $3 \pm 1$\% for the BL. 
For the same component in 4U~1820--303, \cite{dimarco2023} obtained $5.3 \pm 0.2$\%, while in the case of GS~1826-238, because of the intrinsically low polarization
level of the source ($\la 1.3$\%), it was not possible to perform a detailed spectropolarimetric study \citep{capitanio2023}.

\section{Modeling the polarization of the boundary layer around NS}\label{sect_bl_model}

The continuously increasing amount of data coming from  IXPE, as well as future satellites 
 such as the enhanced X-ray Timing and Polarimetry mission \citep[][]{zhang2019}, needs a parallel development of theoretical and numerical simulations, 
which are fundamental for data interpretation.
Several codes for computing polarization in the X-ray band have been developed both in the past and more recently.
The cases ignoring general relativity effects were treated, for instance, by \citet[][hereafter ST85]{st85} and \citet[][hereafter PS96]{ps96},
who considered a slab geometry. Both algorithms 
were based on an iterative procedure, the main difference consisting of the use  of the Thomson (ST85) or Klein-Nishina (PS96) scattering kernel.
Most efforts in the theoretical investigation have so far been focused on the polarization properties of accretion disks
around black holes \citep{schnittman2009, dovciak2011, schnittman2013, abarr2020, taverna2021, Loktev22,podgorny2022}.

Instead, X-ray polarimetry of XRBs hosting an NS has received much less attention, and in this framework
 the seminal paper of \citet[][hereafter LS85]{Lapidus85} deserves to be mentioned, as the authors considered the polarization properties
of X-ray bursters modeled with an accretion disk and a BL approximately at a right angle, both during the burst and the out-of-burst phases.
Despite being a work dating back to the mid-1980s, to our knowledge this is the most detailed treatment of the problem so far.
For our work, it is important to point out that LS85 modeled the BL as an equatorial belt around the NS of geometrical thickness
$\Delta R \ll R_{\rm ns}$ and variable height-scale ratio $H/R_{\rm ns}$.
In particular, LS85 did not consider the direct disk emission and showed that the total PD of radiation directly coming
from the BL plus the fraction intercepted by the disk and reflected toward the observer may achieve 6\% 
for a  viewing angle around $70^{\circ}$. 
The results of LS85 do not consider light bending of radiation from the emission region to the observer frame. 
The problem was instead considered to study spectropolarimetric properties of X-ray pulsars (XRPs), in which the local emission comes
from the hot spots at the magnetic poles \citep[e.g.,][]{pihajoki2018, bogdanov2019}.

The aim of our work is to investigate the polarization properties of the BL using an approach similar to that
used for XRPs, but considering the phase-independent problem in which the BL covers the NS surface
in a belt-like manner.
We adopted a two-phase approach: in Sect. \ref{sect_rte_slab}, we computed the PD and PA of radiation in the BL comoving frame
through the solution of the radiative transfer equation (RTE) for polarized radiation  in
plane-parallel geometry. As long as the radial thickness of the BL is significantly less than the NS radius, this is a good approximation.
Subsequently, we performed angular integration over the BL surface and calculated
the total received signal in the observer frame as a function of inclination  with respect
to the NS spin axis (see Sect. \ref{sect_bl_model}).
We present the numeral results in Sect. \ref{sect_results} and provide a discussion in Sect. \ref{sect_discussion}.
Finally, we draw our conclusions in Sect.  \ref{sect_conclusion}.

\subsection{The radiative transfer equation of polarized radiation in plane-parallel geometry}\label{sect_rte_slab}

We considered plane-parallel geometry and define the total specific intensity as $\Itot=\Il + \Ir,$ where $\Il$ and $\Ir$ are the intensities
with the electric field contained in the meridional plane and perpendicular to it, respectively \citep[hereafter C60]{chandrasekhar1960}.
The meridional plane is defined as the plane formed by the normal to the slab and the 
propagation direction of photons.

The coupled RTEs for an electron-scattering atmosphere in the Thomson limit are \citep{pomraning1973}

\begin{eqnarray}
 \mu\frac{d\Il (\tau, \mu)}{d\tau} &=& -\Il (\tau, \mu)   + \frac{3}{8}\int^{1}_{-1} \mathcal{P}_{1,1}(\mu,\mu^{\prime})\Il(\tau, \mu^{\prime}) {\rm d}\mu^{\prime}  \label{initial_rte_a} \\
&& + \frac{3}{8}\int^{1}_{-1} \mathcal{P}_{1,2}(\mu,\mu^{\prime})\Ir(\tau, \mu^{\prime}) {\rm d}\mu^{\prime}  + S_{\rm l} (\tau, \mu)\nonumber, \\
\mu\frac{d\Ir (\tau, \mu)}{d\tau} &=& -\Ir (\tau, \mu)  + \frac{3}{8}\int^{1}_{-1} \mathcal{P}_{2,1}(\mu,\mu^{\prime})\Il(\tau, \mu^{\prime}) {\rm d}\mu^{\prime} \label{initial_rte_b} \\
&& + \frac{3}{8}\int^{1}_{-1} \mathcal{P}_{2,2}(\mu,\mu^{\prime})\Ir(\mu^{\prime}, \tau) d\mu^{\prime} + S_{\rm r} (\tau, \mu) \nonumber,
\end{eqnarray}

\noindent
where $\mu$ is the cosine of the angle of radiation beam with respect to the normal and $S_{l,r}(\tau)$ is the source term describing the distribution over $\tau$ of the seed
photons. 
We considered initially unpolarized  seed photons; thus, $S_l (\tau)=S_r (\tau)=\frac{1}{2} S (\tau)$.\newline
The Chandrasekhar scattering matrix $\mathcal{P}_{i,j}(\mu,\mu^{\prime})$ is defined as (ST85) follows:

\begin{eqnarray}\label{chandra_matrix}
\mathcal{P}_{1,1}(\mu,\mu^{\prime}) & = & 2 (1-\mu^2) (1-\mu^{\prime 2}),\\
\mathcal{P}_{1,2} (\mu,\mu^{\prime})& = &\mu^{2},  \\
\mathcal{P}_{2,1} (\mu,\mu^{\prime})& = &\mu^{\prime 2},\\
\mathcal{P}_{2,2} (\mu,\mu^{\prime})& = & 1. 
\end{eqnarray}

\noindent
The system of two integro-differential equations -Eqs. (\ref{initial_rte_a}) and (\ref{initial_rte_b})-
for the unknowns $\Il(\tau, \mu), \Ir(\tau, \mu)$ is solved with an iterative procedure with appropriate boundary conditions, similarly
to ST85. 
First, we consider the distribution over $\tau$ of photons which escape without interaction with matter.
Mathematically, this is equivalent to ignoring the scattering terms (i.e., the integrals), leaving
only the source function $S(\tau)$ and the \emph{\emph{absorption}} terms in the right sides of the system.
We note that this is not a true absorption in terms of photon loss due to bound-bound or bound-free
processes, but rather a decrease of the photon intensity in the given direction, $\mu$, as Thomson 
scattering diffuses a fraction of the incoming intensity away from the $\mu$ direction.

For the unscattered radiation, Eqs. (\ref{initial_rte_a}) and  (\ref{initial_rte_b})  then become

\begin{eqnarray}\label{rte_noscattering}
\mu\frac{{\rm d} \Il (\tau, \mu)}{{\rm d}\tau} & = & -\Il (\tau, \mu)+ \frac{S (\tau)}{2},\\
\mu\frac{{\rm d} \Ir (\tau, \mu)}{{\rm d}\tau} & = &- \Ir  (\tau, \mu)+ \frac{S (\tau)}{2}, 
\end{eqnarray}

\noindent
for which the general solution is

\begin{equation}\label{I_zeroscattering}
I^0_{\rm l,r}(\tau, \mu)=C_1 e^{-\tau/\mu}+\frac{1}{2\mu}\int^{\tau}_0 e^{-(\tau-\tau^{\prime})/\mu} S(\tau^{\prime}) {\rm d}\tau^{\prime}.
\end{equation}

\noindent
We define the bottom of the slab at $\tau=0$ and the top at $\tau=2\tau_0$. We note that we conventionally define the total optical depth of the slab as 2$\tau_0$ to keep the same definition of ST85.
The condition of no incident radiation at the base of the slab reads $I_{\rm l,r}(0, \mu)=0$, which implies $C_1=0$. 
The explicit form of Eq. (\ref{I_zeroscattering}) depends on the initial seed photon distribution $S(\tau)$.

We consider two cases here. In the first one, which is labeled Case 1 and is  important for our further discussion,
seed photons have isotropic angular distribution for $\mu \geq 0$ and are located at the base of the atmosphere, namely $S(\tau)=S_0 \delta(\tau)$. We note that with this configuration $S(\tau)=0$ for $\mu < 0$, and 
in this case  Eq. (\ref{I_zeroscattering}) becomes

\begin{eqnarray}
I^0_{\rm l,r} (\tau, \mu) & = &   \frac{S_0}{2\mu} e^{-\tau/\mu}   ~~~~~~ \mu \geq 0,  \label{sol_delta_k0_a}\\
I^0_{\rm l,r} (\tau, \mu) & = &  0               ~~~~~~                 \mu < 0. \label{sol_delta_k0_b} 
\end{eqnarray}

\noindent
For the  second case, labeled Case 2, we consider isotropic seed photons uniformly distributed over the slab, 
 namely $S(\tau)=S_0$, and in this case the solution of the RTE gives

\begin{eqnarray}
I^0_{\rm l,r} (\tau, \mu) & = & \frac{S_0}{2} (1-e^{-\tau/\mu})~~~~~~~~~~~~~\mu \geq 0,  \label{sol_uniform_k0_a}\\
I^0_{\rm l,r} (\tau, \mu)& = & \frac{S_0}{2} \left[1-e^{-(2\tau_0-\tau)/\mu} \right]~~~~~~~~~~~~~\mu < 0.  \label{sol_uniform_k0_b}
\end{eqnarray}

\subsection{Solution for scattering order k >0}

When considering the specific intensity of radiation subjected to $k$ scatterings,
the source term $S(\tau)$ disappears, and the contribution to the specific intensity across the $\mu$ direction comes
from photons of scattering order $I^{k-1}_{\rm l,r}$ diffused into the $\mu$ direction from all
other directions.
This contribution arises from both photons, which remain in  the same polarization mode 
(e.g., $I^{k-1}_{\rm l} \rightarrow I^k_{\rm l}$), as well as those switching from one mode to the other 
(e.g., $I^{k-1}_{\rm r} \rightarrow I^k_{\rm l}$). 

\noindent
The system of Eqs. (\ref{initial_rte_a}) and (\ref{initial_rte_b})  then becomes

\begin{eqnarray}\label{rte_korder}
 \mu\frac{{\rm d} \Ikl (\tau, \mu)}{{\rm d}\tau} & = & -\Ikl (\tau, \mu)+\frac{3}{8}\int^{1}_{-1} \mathcal{P}_{1,1}(\mu,\mu^{\prime})I^{k-1}_{\rm l}(\mu^{\prime}) {\rm d}\mu^{\prime}\\
&& +\frac{3}{8}\int^{1}_{-1} \mathcal{P}_{1,2}(\mu,\mu^{\prime})I^{k-1}_{\rm r}(\mu^{\prime}) {\rm d}\mu^{\prime},\nonumber \\
\mu\frac{{\rm d}\Ikr (\tau, \mu)}{{\rm d}\tau} & = & -\Ikr (\tau, \mu) +\frac{3}{8}\int^{1}_{-1} \mathcal{P}_{2,1}(\mu,\mu^{\prime})I^{k-1}_{\rm l}(\mu^{\prime}) {\rm d}\mu^{\prime} \\
&& +\frac{3}{8}\int^{1}_{-1} \mathcal{P}_{2,2}(\mu,\mu^{\prime})I^{k-1}_{\rm r}(\mu^{\prime}) {\rm d}\mu^{\prime}. \nonumber
\end{eqnarray}

\begin{figure}
\includegraphics[width=0.5\textwidth]{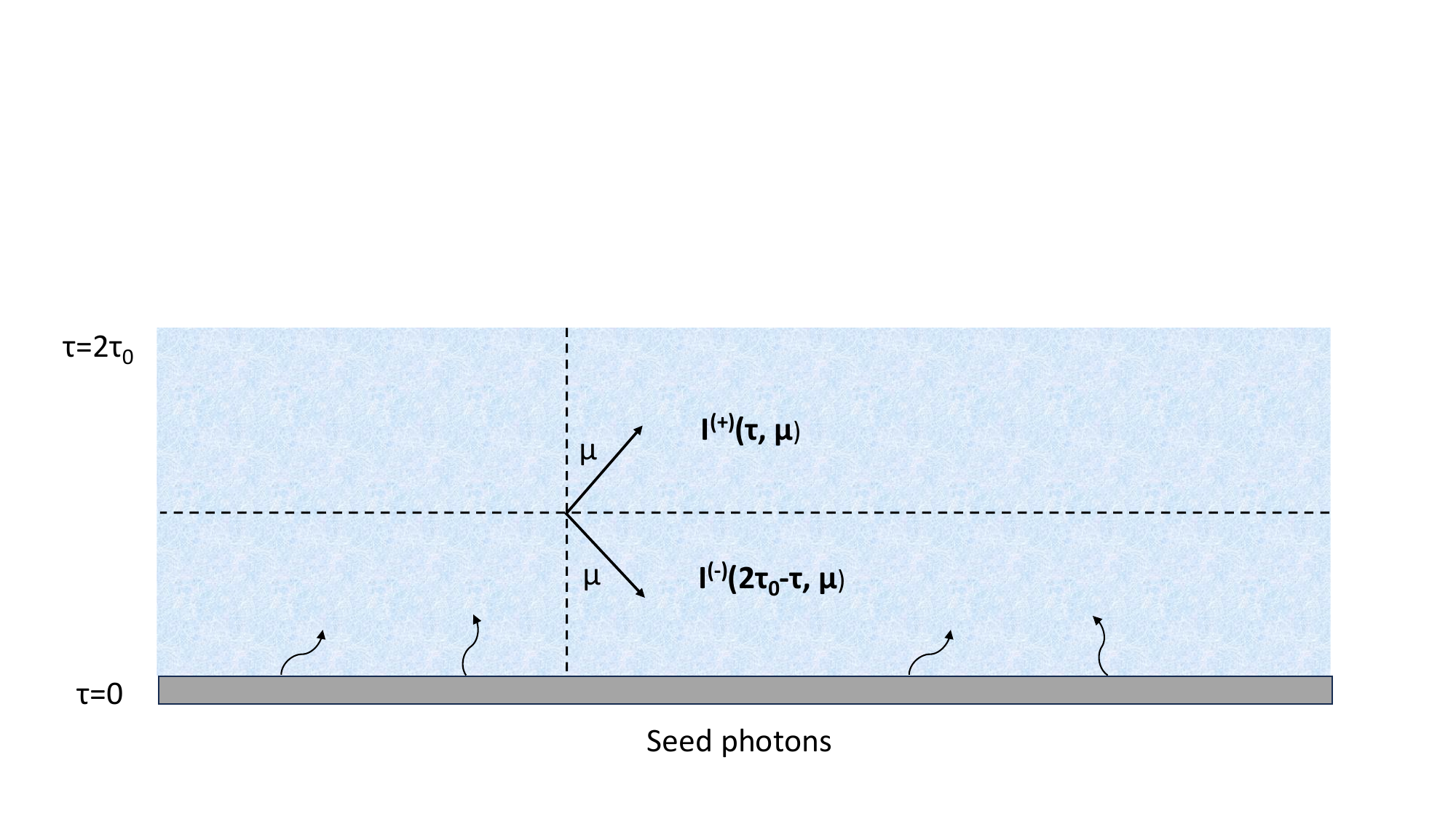}
\caption{Sketch of radiative transfer problem in plane-parallel geometry.
Seed photons at the base of the slab are shown here. At any location $\tau$, a positive angle $\mu$ relative to the slab normal can be defined 
for an upward and downward intensity component, respectively.}
\label{slab_geometry}
\end{figure}

\noindent
As the angular integration is performed over the domain  $\mu^{\prime} \in [-1:1]$, we define the \emph{\emph{upstream}} intensity $\Iup(\tau, \mu)$ as the one directed toward the top  layer of the slab, and \emph{\emph{downstream}} intensity $\Iminus (\tau, \mu)$ as that directed toward the bottom layer.
In this way, the variable $\mu$ is positively defined for both $\Iup(\tau, \mu)$ and $\Iminus (\tau, \mu)$, 
from bottom to top in the first case, and the opposite in the second one (see Fig.\ref{slab_geometry}).

\noindent
Considering, for example, the $l$-mode, the integral terms can thus be written as 

\begin{eqnarray}
\int^{1}_{-1} \mathcal{P}_{i,j}(\mu,\mu^{\prime})I^{k-1}_{\rm l}(\tau,\mu^{\prime}) {\rm d}\mu^{\prime}  =  \\
\int^{1}_{0} \mathcal{P}_{i,j}(\mu,\mu^{\prime}) \left[\Iuplkminusone(\tau,\mu^{\prime}) + \Iminuslkminusone(2\tau_0-\tau,\mu^{\prime}) \right] {\rm d}\mu^{\prime}.  \nonumber
\end{eqnarray}

\noindent
Given that the intensities $I_{\rm r,l}(\tau, \mu)$ are now expressed as the contribution of the
upstream and downstream factors, the coupled RTEs switch from two to \emph{\emph{four}} integro-differential equations 
according to

\begin{eqnarray}
\mu\frac{{\rm d}\Iuplk (\tau, \mu)}{{\rm d}\tau}=-\Iuplk (\tau, \mu)   \label{num_system_1} \\
+\frac{3}{8}\int^{1}_{0} \mathcal{P}_{1,1}(\mu,\mu^{\prime}) \left[\Iuplkminusone(\tau, \mu^{\prime})  + \Iminuslkminusone (2\tau_0-\tau, \mu^{\prime})\right] {\rm d}\mu^{\prime} &  &\nonumber \\
+\frac{3}{8}\int^{1}_{0} \mathcal{P}_{1,2}(\mu,\mu^{\prime}) \left[\Iuprkminusone(\tau, \mu^{\prime})  +\Iminusrkminusone(2\tau_0-\tau, \mu^{\prime})\right] {\rm d}\mu^{\prime}, &  & \nonumber \\
\mu\frac{{\rm d}\Iuprk (\tau, \mu)}{{\rm d}\tau}=-\Iuprk (\tau, \mu) \label{num_system_2}\\
+\frac{3}{8}\int^{1}_{0} \mathcal{P}_{2,1}(\mu,\mu^{\prime}) \left[\Iuplkminusone(\tau, \mu^{\prime})  +\Iminusl(2\tau_0-\tau, \mu^{\prime})\right] {\rm d}\mu^{\prime} \nonumber  &  & \nonumber \\
+\frac{3}{8}\int^{1}_{0} \mathcal{P}_{2,2}(\mu,\mu^{\prime}) \left[\Iuprkminusone(\tau, \mu^{\prime})  +\Iminusrkminusone(2\tau_0-\tau, \mu^{\prime})\right] {\rm d}\mu^{\prime}, & & \nonumber \\
\mu\frac{{\rm d}\Iminuslk(\tau, \mu)}{{\rm d}\tau}=-\Iminuslk (\tau, \mu) \label{num_system_3}\\
+\frac{3}{8}\int^{1}_{0} \mathcal{P}_{1,1}(\mu,\mu^{\prime}) \left[\Iuplkminusone( 2\tau_0 -\tau ,\mu^{\prime})  + \Iminuslkminusone  (\tau, \mu^{\prime})\right] {\rm d}\mu^{\prime}\nonumber & & \nonumber \\
+\frac{3}{8}\int^{1}_{0} \mathcal{P}_{1,2}(\mu,\mu^{\prime}) \left[\Iuprkminusone(2\tau_0-\tau, \mu^{\prime})  + \Iminusrkminusone (\tau, \mu^{\prime})\right] {\rm d}\mu^{\prime}, & & \nonumber \\
\mu\frac{{\rm d}\Iminusrk(\tau, \mu)}{{\rm d}\tau}=-\Iminusrk(\tau, \mu) \label{num_system_4}\\
+\frac{3}{8}\int^{1}_{0} \mathcal{P}_{2,1}(\mu,\mu^{\prime}) \left[\Iuplkminusone(2\tau_0 - \tau, \mu^{\prime})  +\Iminuslkminusone (\tau, \mu^{\prime})\right] {\rm d}\mu^{\prime} \nonumber & & \nonumber\\
+\frac{3}{8}\int^{1}_{0} \mathcal{P}_{2,2}(\mu,\mu^{\prime}) \left[\Iuprkminusone(2\tau_0 - \tau, \mu^{\prime})  + \Iminusrkminusone(\tau, \mu^{\prime})\right] {\rm d}\mu^{\prime}.\nonumber
\end{eqnarray}

\noindent
The first two equations of the system solve for the upstream intensities $\Iuplk$ and $\Iuprk$ with $\tau$ computed from bottom to top, while the other two solve for the downstream components $\Iminuslk$ and $\Iminusrk$
with $\tau$ computed from top to bottom. Once defined the seed photons distribution $S(\tau)$, the algorithm first seeks for the solution 
$I^0_{l,r} (\tau, \mu)$ as defined in Eqs. (\ref{sol_delta_k0_a})-(\ref{sol_delta_k0_b}) or (\ref{sol_uniform_k0_a})-(\ref{sol_uniform_k0_b}), and then iteratively integrates over $\tau$ equations 
up to a user-defined maximum number of scatterings $k^{\rm max}$.

The values of the specific intensity in Eqs. (\ref{num_system_1}) to (\ref{num_system_4}) are computed
at fixed values of $\mui,$ which are then  preliminarily defined over a given grid, and 
 the angular integration over $\mu^{\prime}$  is performed using
the Gauss-Legendre (GL) quadrature rule. We remind the reader that GL quadrature is used for computing the
integral of functions bounded over the domain
$x \in [a:b]$ according to

\begin{equation}\label{gauss_legendre}
\int^b_a f(x) {\rm d}x \approx \sum^{N-1}_{i=0} f(x_i) w_i,
\end{equation}

\noindent
where $a < x_i < b$ are the roots of the $N^{\rm th}$ Legendre polynomial, and $w(x_i)$ are the quadrature weights. 
Of course, the higher the value of $N$, the greater the precision in approximating the true integral value.
We checked that for $N\ga 20$ there are no further appreciable changes in numerical results, so we fixed $N=30$.
We eventually found that it was convenient to compute the solution at the $\mui$ values coincident with the roots of the GL quadrature.
The numerical integration over $\tau$ was performed using an explicit embedded Runge-Kutta Prince-Dormand method
with adaptive step-size control provided by the Gnu Scientific Library\footnote{https://www.ict.inaf.it/gitlab/polarization/slab/-/releases}.

By definition, the iterative method provides the solution for each scattering order $\Ikrl(\tau, \mu)$.
We are interested in the following quantities at the top of the slab ($\tau=2\tau_0$), where it is
assumed that an observer is located:
angular distribution of the total specific intensity (limb-darkening
law); angular distribution of the total polarization.

Defining the total intensities of the two polarization modes $(r,l)$ at the top of the slab
by summing over all the orders of scattering, we obtain

\begin{equation}\label{Irl_tot}
\Itotrlup(\mu) =\sum^n_{k=0} I^{(+) k}_{\rm {r,l}}(\mu, 2\tau_0),
\end{equation}

\noindent
and the total polarization comes directly to 

\begin{equation}\label{pdeg_tot}
P^{\rm tot}(\mu)=\frac{\Itotrup(\mu) - \Itotlup(\mu)}{\Itotrup(\mu) + \Itotlup(\mu)}.
\end{equation}

\noindent
Dropping the "tot" superscript for clarity, one may see that $P(\mu)$ can be positive or negative. As in planar and
spherical configurations, the Stokes parameter $U=0$; here, $P=Q$.
Keeping in mind the definition of the PA,

\begin{equation}\label{pa_tot}
\chi=\frac{1}{2}{\rm Atan} \frac{U}{Q},
\end{equation}

\noindent
it results that for $Q > 0$, $\chi$=0; otherwise, $\chi=\pi/2$.
The PA is then parallel or perpendicular to the slab plane for positive and negative values of the PD,
respectively.

We also considered the percentage of the flux emerging from the top of the slab, where the comoving observer is located.
From the definition of flux and using the GL quadrature rule, one obtains

\begin{equation}
\Fplus=\sum^{N-1}_{i=0} \Itotrlup (\mu_i) \mu_i w(\mu_i),
\label{eq_fplus}
\end{equation}

\begin{equation}
\Fminus=\sum^{N-1}_{i=0} \Itotrldown (\mu_i) \mu_i w(\mu_i),
\label{eq_fminus}
\end{equation}

\begin{equation}
\epsilon (2\tau_0)=\frac{\Fplus}{\Fplus+\Fminus},
\label{eq_frac_fplus}
\end{equation}

\noindent
where $\Fplus$ and $\Fminus$ are the upstream ($\tau=2 \tau_0$) and downstream fluxes ($\tau=0$), respectively.

The parameter $\epsilon  (2\tau_0)$ is by definition equal to 0.5 for seed photons distribution symmetric with respect to the center of the slab (as in ST85), while it depends on the  optical depth for seed photons located at the base.
 
\subsection{Polarimetry of the boundary layer}

\begin{figure}
   \includegraphics[width=0.5\textwidth]{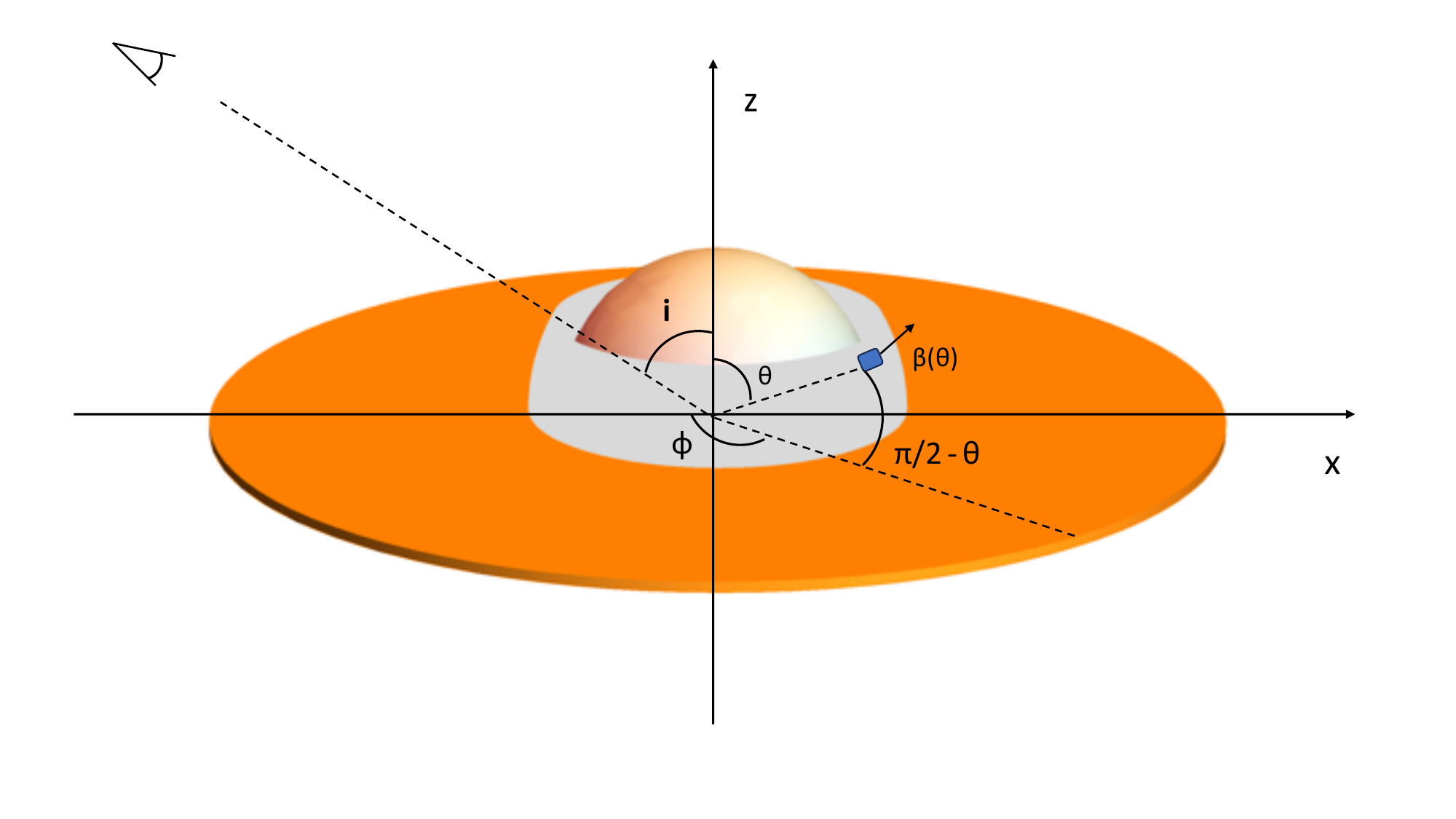}
   \caption{Proposed geometry for soft state of NS-LMXBs; the disk extends very close to the NS surface, and then a BL forms with $\Delta R \ll \Rns$. The BB photons from the NS photosphere are located at the base 
   of the BL, which extends up to some latitude $\thetamax$ computed from the orbital plane, and has a given temperature $\kte$ and radial optical depth 2$\tau_0$. Only photons in the northern hemisphere ($\theta < \pi/2$) reach the observer. The total signal is obtained by integrating the contribution from each elementary area $\dsigmaprime$ shown by the blue box over the
BL surface.}
   \label{bl_3dgeom}
    \end{figure}

The PD and PA in Eqs. (\ref{pdeg_tot}) and (\ref{pa_tot}) are defined
at the top of the slab, which is identified as the surface of the BL (i.e.,  comoving frame).
As already mentioned in Sect.  \ref{sect_introduction}, the mathematical formalism for passing from the comoving to the observer frame has been developed  to investigate spectropolarimetric properties of accreting XRPs.
We consider the method defined in \citet[][hereafter P20]{poutanen2020}, to which we refer 
the reader for details.
The most significant difference between our work and results obtained for XRPs is the change in 
the contribution to the total observed PD and PA from just two small spots at the magnetic poles of the NS
to an equatorial belt covering the NS surface up to a given latitude $\thetamax$.
This can be achieved thanks to the additive property of the Stokes parameters, so the 
computed quantities can be derived as the contribution from each elementary area d$\Sigma^{\prime}$ of the BL surface
integrated over the domain $\theta \in [\pi/2 - \thetamax: \pi/2]$ and $\phi \in [0:2\pi]$ 
(see Fig. \ref{bl_3dgeom} and Eqs. [17]-[21] in P20).
The proper area element using angular coordinates is

\begin{equation}
 \dsigmaprime= \gamma (\theta) R {\rm sin}\theta ~{\rm d}\theta~{\rm d}\phi,
\end{equation}

\noindent
where $\gamma(\theta)$ is the Lorentz factor of the matter velocity at given latitude with

\begin{equation} \label{beta_value}
    \beta(\theta)=\frac{2 \pi \Rbl}{c} \frac{\blfreq} {\sqrt{1-\eta}} {\rm sin} \theta.
\end{equation}

\noindent
The impact parameter in Eq. (\ref{beta_value}) is $\eta=\Rg/\Rbl$, where $\Rbl \ga \Rns$, and for which we set the value 
$\eta=1/2.5$, while we consider a BL co-rotating with the NS at frequency  $\blfreq=500$ Hz.

For each elementary area ${\rm d}\Sigma^{\prime}$ at coordinates $\theta$ and $\phi$ on the BL surface, one defines the Stokes vector in the observer frame as

\begin{equation}
{\bf \mathcal{S}}= \Itot {\rm d} \Omega [1,  P~{\rm cos}~2\chi, P~{\rm sin}~2\chi].
\end{equation}For the relation between ${\rm d}\Omega$ and ${\rm d}\Sigma^{\prime}$ as well as $\Itot$ and $P$ between the observer and 
comoving frame, we refer the reader to Eqs. (17)-(20) of P20, while $\chi$ is given in Eq. (58) of the same paper and is the PA with respect to the projection of the NS spin onto the sky.



By defining $\Stot$ as the Stokes vector obtained integrating ${\bf \mathcal{S}}$ over d$\Omega$,
the total PD and PA are then derived using the standard definitions, with $Q$ and  $U$ as the second and third
component of $\Stot$, respectively (see Eq. [\ref{pa_tot}]).
It is important to note that the upper boundary for the polar angle is $\theta=\pi/2$, because only radiation from the BL northern hemisphere is received, the southern part being totally intercepted by the inner disk
(and vice versa for an observer located in the opposite position).

While performing two-dimensional integration over the BL surface, the comoving frame angular dependence of the specific intensity and PD are obtained
by a linear interpolation of the array of values stored in a text file and preliminarily computed for plane-parallel
geometry. Numerical results were obtained with a C code optimized with a 
 third-party  adaptive integrator\footnote{https://www.ict.inaf.it/gitlab/polarization/boundary\_layer/-/releases}.

\section{Results}\label{sect_results}

We first discuss the results obtained from the solution of the RTE in plane-parallel geometry 
(see Eqs. [\ref{initial_rte_a}] and [\ref{initial_rte_b}]).
The solutions presented by ST85 considered three different configurations for the initial
seed photons distribution, namely \emph{i)} sources at the center of the slab, \emph{ii)} uniform photon distribution,
 \emph{iii)} and photons at the boundaries. 
 These configurations are by definition totally symmetric with respect to the center of the slab, and there
 is no difference while considering physical quantities in the upward or downward direction.
 We consider a uniform seed photon distribution as well, but unlike the ST85 case we set up a geometry
 where seed photons are located at the bottom of the slab. In this case, the upward and downward symmetry breaks and
 different spectropolarimetric values are obtained at $\tau=0$ or $\tau=2\tau_0$. 

We focus on the results at the top of the slab, where we assume that the comoving observer is located.
A configuration with a plane-parallel electron-scattering atmosphere located above a source of photons
can be representative of a geometrically thin BL partially covering the surface of the NS, with $\Delta R \ll \Rns$, or the atmosphere
above an accretion disk. 
The latter case was also recently investigated by \cite{taverna2021}
for both the cases of pure electron scattering and a combination of scattering plus absorption; although,
the authors reported their results only in the rest frame of the atmosphere.
The case of uniform seed photons distribution, on the other hand, is in our opinion more suitable
to be considered if they originate from free-free emission processes \citep{narayan95}.
However, we present results concerning observational differences between the two configurations here.

\begin{figure} 
   \includegraphics[height=10cm, width=13cm]{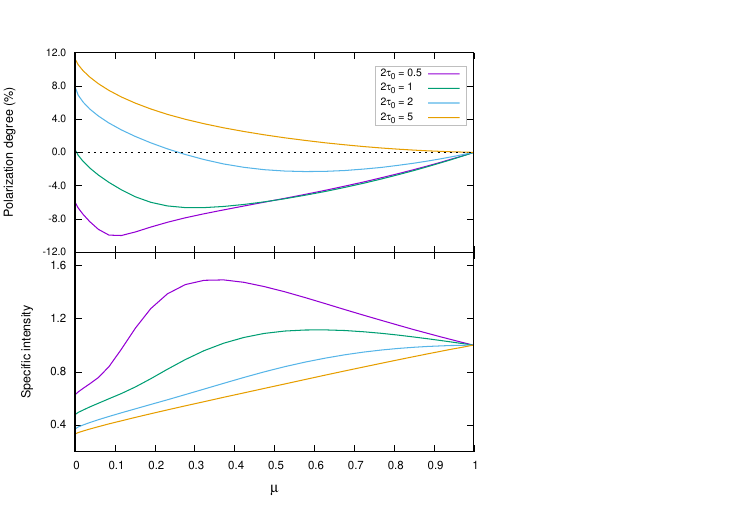}
     \caption{Angular distribution of total PD (\emph{upper panel}) and normalized specific intensity (\emph{lower panel}) emerging from the upper layer of the plane-parallel atmosphere and different values of the optical depth. Seed photons are located at the base of the slab (Case 1). For positive values of the PD, the PA is coplanar to the slab; for negative values it is parallel to the normal.}
  \label{fig_pdegintensity_seed3}   
\end{figure}

\begin{figure} 
   \includegraphics[height=10cm, width=13cm]{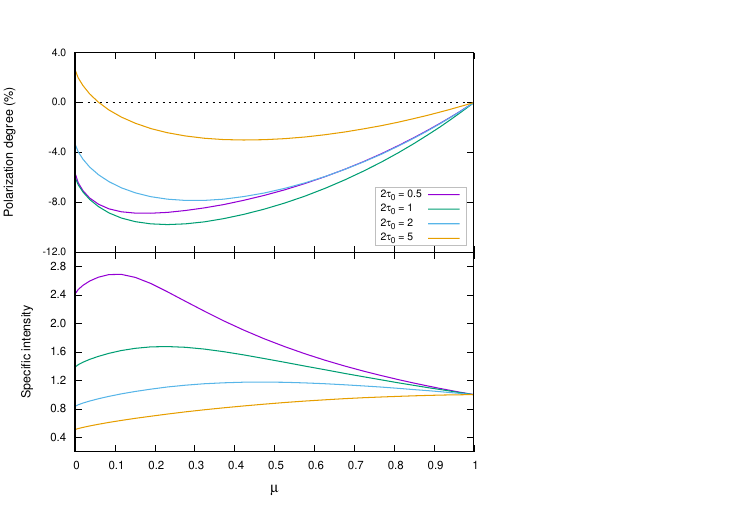}
     \caption{Same as Fig. \ref{fig_pdegintensity_seed3}, but for seed photons uniformly distributed over $\tau$ (Case 2).}
  \label{fig_pdegintensity_seed2}   
\end{figure}

In Fig. \ref{fig_pdegintensity_seed3}, we report the angular behavior of the total specific intensity 
and  PD for seed photons located at the base of the slab (Case 1) and for different values of the optical
depth. The specific intensity values are normalized to the case $\mu=1$.
When $2\tau_0 \la 1$, the PD is always negative, which implies a PA aligned with the slab normal.
For the intermediate value $2\tau_0=2$, PA swapping occurs around $\mu \sim$ 0.3, while for  $2\tau_0=5$ the solution
 becomes virtually indistinguishable from the Chandrasekhar limit.
When seed photons are uniformly distributed, on the other hand, except for a small interval at $\mu \la 0.1$ for $2\tau_0=5$, PD is always negative, as shown in Fig. \ref{fig_pdegintensity_seed2}.
For low optical depths, with both configurations the total intensity $\Itot (\mu)$ does not show monotonic behavior (see also Fig. 4 of ST85); however,
it should be kept in mind that the observed flux at large distances is $\ftot (\mu) \propto \Itot (\mu) \mu$, and in the case
of an accretion disk atmosphere this mitigates the first-glance impression that at some more edge-on viewing angles the flux
is higher than the face-on case (neglecting relativistic light bending effects).
 

\begin{figure}[h]
\centering
   \includegraphics[height=12cm, width=9cm]{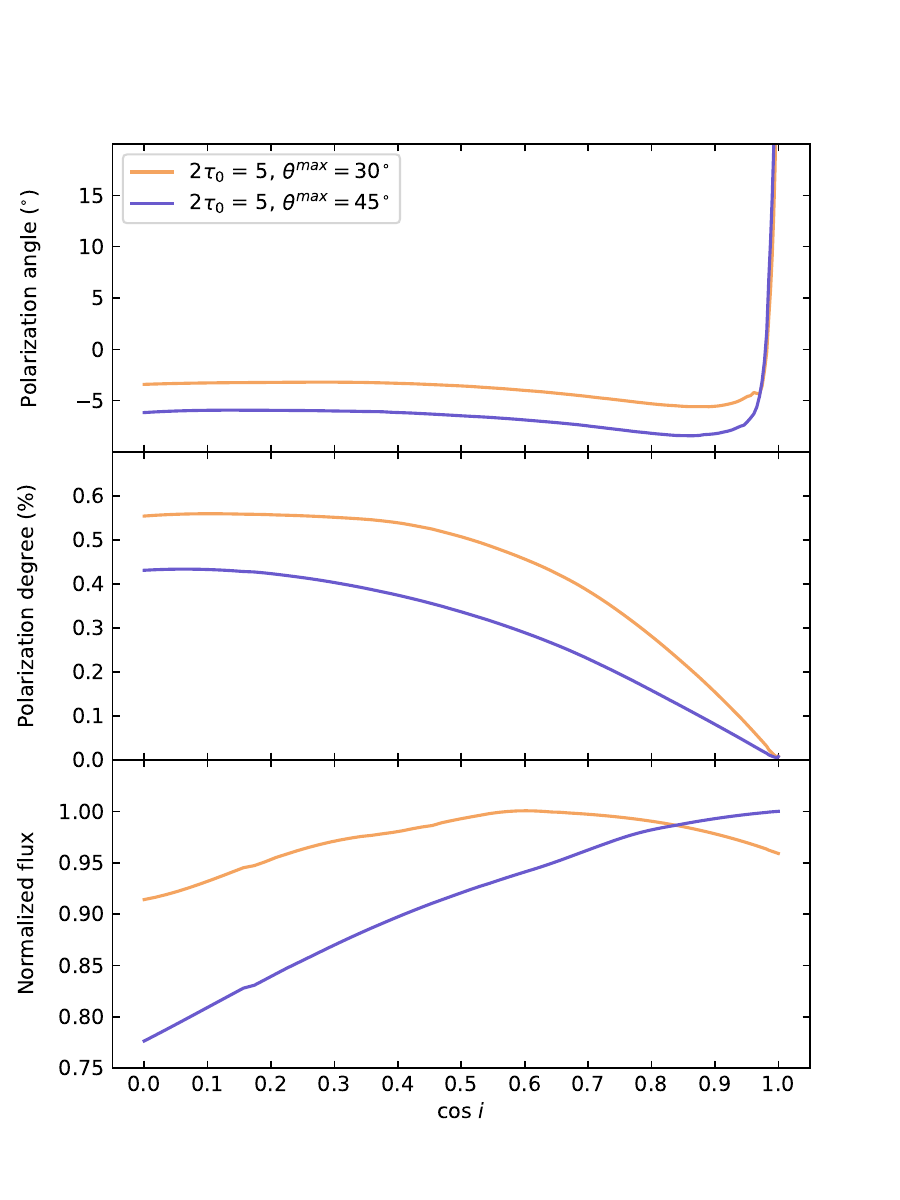}
     \caption{Total PA (\emph{upper panel}), PD (\emph{middle panel}), and normalized intensity (\emph{lower panel})
     of the BL as a function of the viewing angle with seed photons at the base of the slab (Case 1) 
     for an optical depth of $2\tau_0$=5, spin frequency of $\blfreq=500$ Hz, 
     and two values of the BL latitude. The PA is measured with respect to the projection on the sky of the NS spin axis. }
     \label{fig_bl_case1_tau5}
\end{figure}

\begin{figure}[h]
\centering
   \includegraphics[height=12cm, width=9cm]{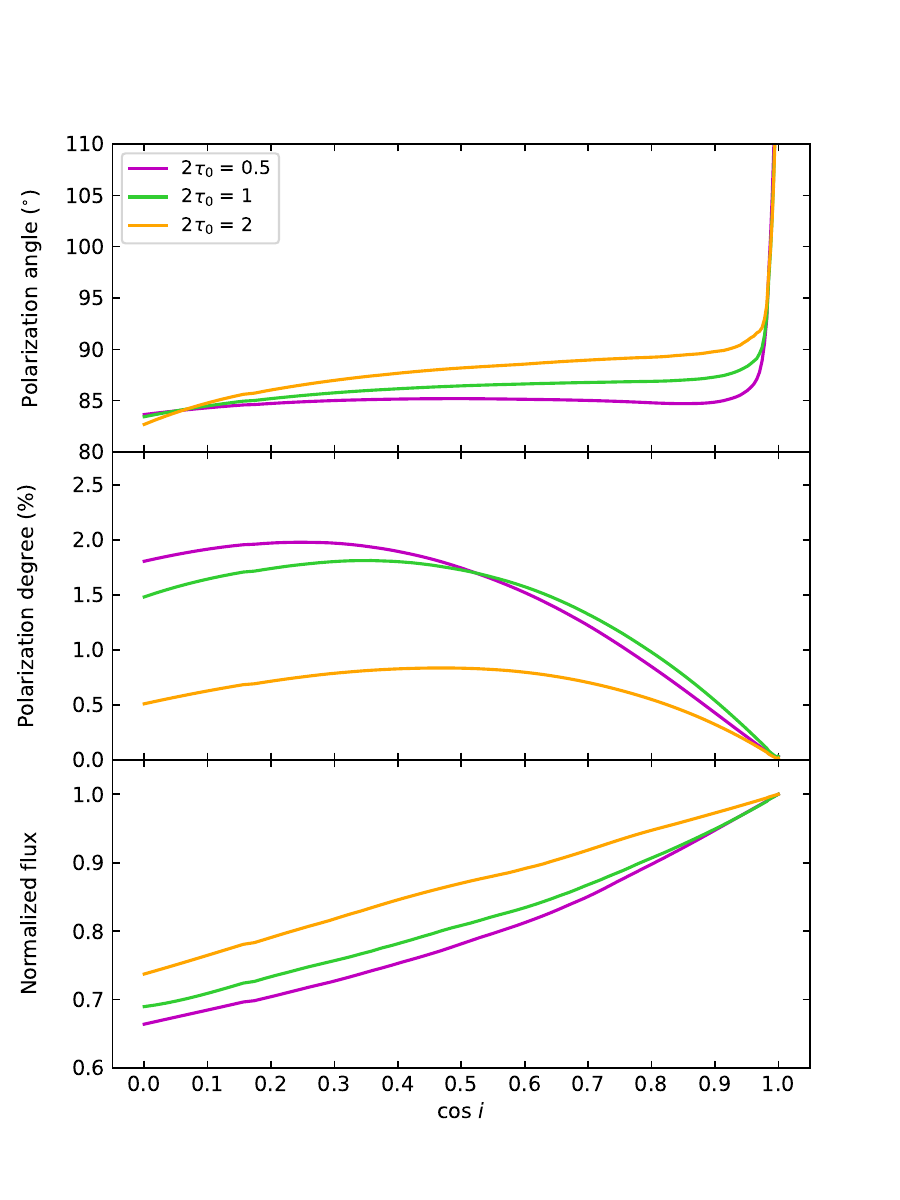}
     \caption{Same as in Fig. \ref{fig_bl_case1_tau5}, but for different values of the optical depth and fixed BL latitude $\thetamax=45^{\circ}$ and spin $\blfreq=500$ Hz.}
      \label{fig_bl_case1_smalltau}
\end{figure}

\begin{figure}[h]
\centering
   \includegraphics[height=12cm, width=9cm]{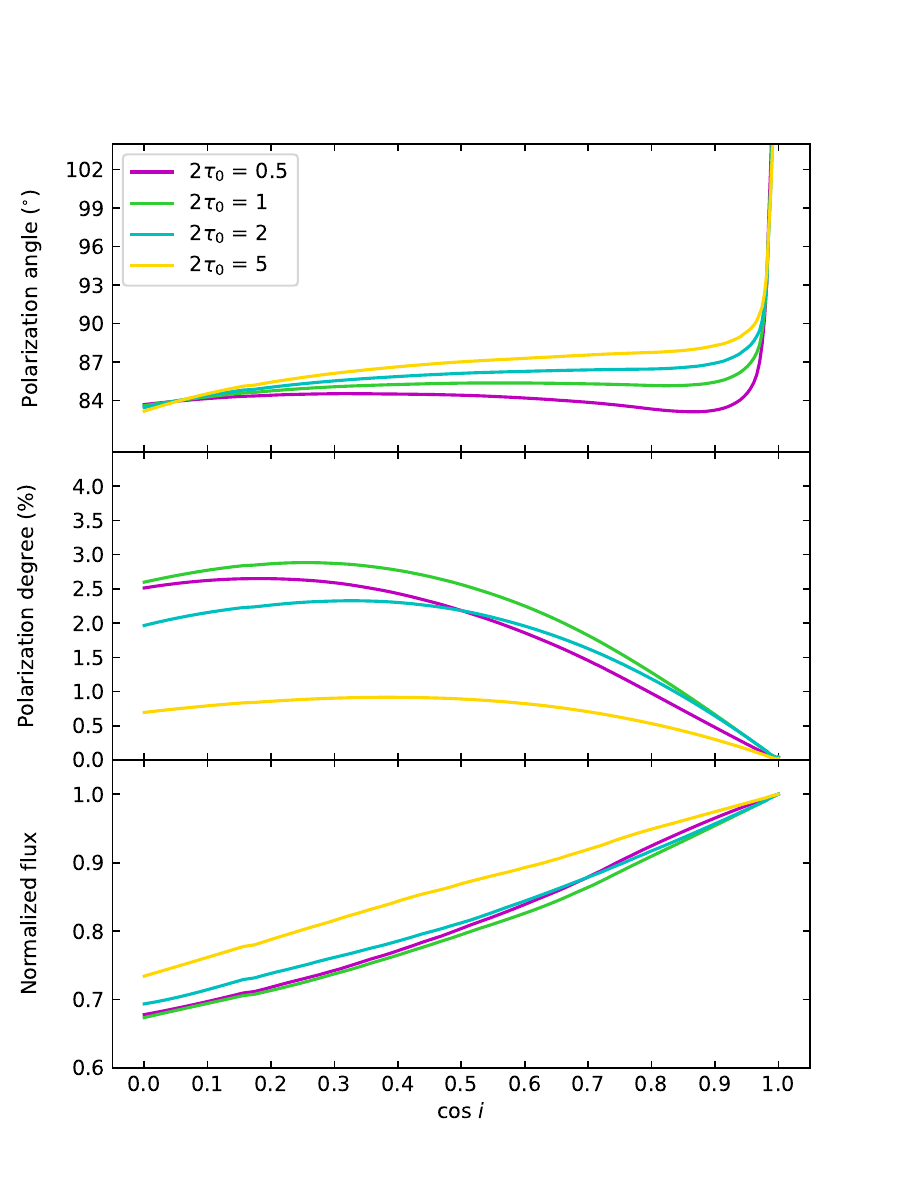}
     \caption{Same as Fig. \ref{fig_bl_case1_tau5}, but with seed photons uniformly distributed (Case 2) and different values of the optical depth of the  BL, with a latitude of $\thetamax=45^{\circ}$ and spin of $\blfreq=500$ Hz.}
     \label{fig_bs_case2_alltau}
\end{figure}

With spectropolarimetric quantities at hand in the comoving frame (i.e., the top of the slab), 
we computed the polarimetric parameters as well as the flux in the observer frame as a function
of the viewing angle.
The results with high optical depth deserve particular attention because this is a typical value of the 
Comptonization component of NS-LMXBs in the soft state (see references in Sect. \ref{sect_introduction}).

We considered two maximum latitudes of the BL, namely $\thetamax=45^{\circ}$ and $\thetamax=30^{\circ}$, with results for Case 1 and optical depth $2\tau_0$=5 shown
in Fig. \ref{fig_bl_case1_tau5}.
As expected, the PD for $\thetamax=45^{\circ}$ is lower than that for $\thetamax=30^{\circ}$, 
as a consequence of the qualitative fact that a less extended BL is intrinsically "spherically less symmetric" 
than a more extended one --actually
a BL covering the whole NS surface would have, by definition, null polarization.
It is also interesting to note that when $\thetamax=45^{\circ}$, the observed flux monotonically increases  
by about 30\%,  while passing from an edge-on to a face-on view of the system, while for $\thetamax=30^{\circ}$ it achieves its maximum value at $i \sim 50^{\circ}$.
The BL properties for lower values of $2\tau_0$ are instead shown in Fig. \ref{fig_bl_case1_smalltau} for
 $\thetamax=45^{\circ}$.
 Here, a PD value of less than 1\% and thus comparable to that of Fig. \ref{fig_bl_case1_tau5} is obtained
 for $2\tau_0$=2, but the PA is $90^{\circ}$ apart, thus making the two cases easily distinguishable
 from the observational point of view. The angular dependence of the observed flux is, on the other hand, qualitatively 
 similar to the case of higher optical depth and $\thetamax=45^{\circ}$.
 \noindent
 
We then considered the configuration with seed photons uniformly distributed over the slab (Case 2), with results
presented in Fig. \ref{fig_bs_case2_alltau}. 
Here, the PD for $2\tau_0$=5 is comparable to that of Case 1, but again the PA is rotated by $90^{\circ}$, 
avoiding degeneracy in the geometrical setup of the input radiation. The flux hence increases monotonically
while progressively decreasing the observer viewing angle.
\noindent
A shared property in  all cases is that the PA is not strictly aligned to the NS spin, or at a right angle to it, due to polarization
plane rotation being the effect of the BL angular velocity.

As we assumed a BL corotating with the NS, the effect depends on the  spin value that  we set to 
$\blfreq$=500 Hz. Lower values of $\blfreq$ of course decrease the deviation of PA from the two
orthogonal directions, leaving a shift of $\la 2^{\circ}$ due only to the light-bending effect for the case
of nonrotating NS/BL.
Regardless of PA rotation being the result of special and general relativistic effects,  \emph{\emph{the corresponding PD 
for high optical depth of the BL is} $\la 0.5$\%}.

\begin{figure}[h]
\includegraphics[width=9cm]{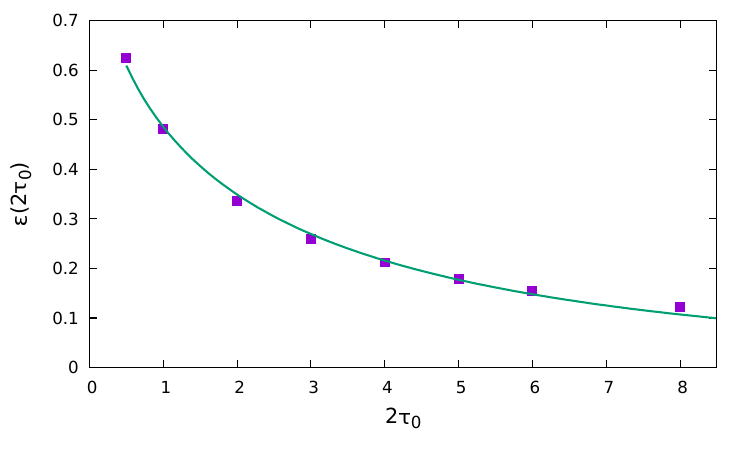}
\caption{Fraction of flux emerging from the top of the slab in the comoving frame as a function of the optical depth (see Eqs. [\ref{eq_fplus}]-[\ref{eq_frac_fplus}]) when seed photons are located at the base.}
\label{fig_ftop_vs_tau}
\end{figure}

\section{Discussion}\label{sect_discussion}

\noindent
The behavior of the physical quantities presented in this work must be considered in their integral form; namely, we consider 
the angular dependence of intensity and  PD in the NS comoving frame (i.e., the top layer of the slab) obtained by summing
over all orders of scatterings (see Eqs. [\ref{Irl_tot}] and [\ref{pdeg_tot}]).
Our choice  is motivated by two main reasons, which we explain below.
The first one is that linking the photon energy of the emerging spectrum to the number of scatterings experienced
in the medium is not as trivial as it could appear: for an electron thermal distribution, the energy of a photon 
that escapes the medium after $k$ scattering is \citep{rl79}

\begin{equation}\label{energy_gain}
E_{k} \approx E_0 \left(1+ \frac{4 \kte}{m_{\rm e} c^2}\right)^k,
\end{equation}

\noindent
where $E_0$ is the initial energy. 
This is true as long as  $E_0 \la E_{\rm k} \ll 4 \kte$; otherwise, the recoil effect is to be taken into account. 
For the typical electron temperature of NS-LMXBs in the soft state ($\kte \sim$ 3--5 keV), the validity
of Eq. (\ref{energy_gain}) fails in the energy interval that partially overlaps with the IXPE 
bandwidth (2-8 keV) so that the identification $I_{\rm l,r}^k(\mu) \equiv I_{\rm l,r}(E_k)$ appears too rough
an approximation.
We note that ST85 presented results both for the energy-integrated spectrum, as in Eq. (\ref{Irl_tot}), and
for what they call qualitatively "hard radiation spectrum", namely for $I_{\rm l,r}^k(\mu)$ with $k > \overline{k}$, the average number of scatterings. When discussing the relation between the number of scatterings and the energy,
they were only able to quantify this in the power-law-like regime of the emerging spectrum.
It is also worth pointing out that Eq. (\ref{energy_gain}) holds \emph{\emph{on average}},
namely there may be photons that are subjected to $k>0$ scattering and leave the medium
with energy lower than the initial one.
The contributions of photons leaving the medium with lower or higher energies than input one is dictated
by the shape of Green's function (GF), for which $I(E) \propto E^{\alpha+3}$ for $E < E_0$ and 
$I(E) \propto E^{-\alpha}$ (up to $E\sim \kte$) for $E > E_0$, and where the spectral index $\alpha$ 
depends on the temperature and optical depth \citep[][hereafter ST80]{st80}.
The red wing of the GF is steeper than the blue wing; however, for an electron temperature
of the order of a few kilo-electron-volts, the power-law regime of the GF (defined by Eq. [\ref{energy_gain}])
is confined to a narrow band.
In turn, the exact form of $I(E)$ and PD$(E)$ requires either Monte Carlo methods or an explicit dependence
on energy of the RTE (see Eqs. [\ref{initial_rte_a}] and [\ref{initial_rte_b}]).
The second reason is the fact that despite its challenging results, the data statistics of IXPE observations
of NS-LMXBs do not allow thorough investigations of detailed energy-dependent trends for PD and PA 
for each spectral component.
When the behavior of PD and PA as a function of energy is presented, it indeed refers to a model-independent data analysis
(see references in Sect.  \ref{sect_introduction}), in which the polarimetric quantities are the
results of contribution from all components of the systems (Comptonization region, accretion disk, as
well as disk reflection).
For model-dependent spectropolarimetric analysis instead, the PD and PA of the single
components are given in the whole IXPE band (2--8 keV).
We thus claim that presenting our energy-integrated theoretical results (i.e., summed over all scattering orders $k$) 
is sufficient to provide theoretical support to IXPE model-dependent data interpretation.

One potential objection to our results could arise from the fact that we used the RTE of polarized radiation in the 
Thomson approximation (i.e., the Chandrasekhar scattering matrix) to compute the 
polarization properties of radiation emerging from the BL (which we identify as the region responsible for the Comptonization spectrum of NS-LMXBs).
The difference from an energy-dependent treatment of the electron scattering cross-section is marginal in 
this context, however.
Indeed, the electron temperature of the Comptonization medium in the NS-LMXBs soft state never exceeds $\kte \sim$ 3--5 keV
\citep[e.g.,][]{done2007}, and at this temperature the Klein-Nishina corrections are marginal.

The same considerations can be of course applied to polarimetry as well.
The post-scattering PD of unpolarized and polarized radiation is \citep{matt1996}

\begin{equation}
    \Pi_{\rm U}=\frac{1-{\rm cos^2} \Theta}{1+{\rm cos^2} \Theta -2 + \eta + 1/\eta},
\end{equation}

\begin{equation}
    \Pi_{\rm P}=2 \frac{1- {\rm sin^2} \Theta ~{\rm cos^2} \Phi}{\eta+ \eta^{-1}-2 {\rm sin^2} \Theta ~{\rm cos^2} \Phi},
\end{equation}

\noindent
where $\eta$ is the ratio of the photon energies after and before scattering in the electron rest frame.
For $\kte \la 5$ keV, it is easy to see that $\eta \sim 1$, and  $\Pi_{\rm U}$-$\Pi_{\rm P}$ have a tiny dependence on energy.
\noindent
Considering different optical depths of the slab and two different distributions of the seed photons, we find the 
following main results.

\begin{itemize}
    \item [-] 
    A PA almost aligned with the NS spin can be obtained only for optically thick configurations and \emph{\emph{seed photons located at the base of the medium}}. 
    Nevertheless, with this physical and geometrical setup, the total PD of the BL is $\la$ 0.5\%.
    Up to now, measurements of absolute PA alignment have been possible only for the Z sources \cygx \citep{farinelli2023}
    and \scox; while in the former case the PA resulted to be parallel with the direction of the radio jet,
    in the second a polarization rotation was found by \cite{lamonaca2023} with respect to previous observations
    performed with PolarLight in which the PA was approximatively aligned with the radio jet as well 
    \citep{long2022}. 
    \item[-] 
    For PA in the direction of the NS spin axis, the value is slightly rotated by $\sim 4^{\circ}$--$7^{\circ}$ for two fiducial values of the BL latitude (30$^{\circ}$ and 45$^{\circ}$, respectively) and BL rotational frequency $\blfreq=500$ Hz as a result of both special and 
     general relativistic effects. For the extreme case of a nonrotating NS/BL system, we checked a deviation due to light bending effect 
      $\la 2^{\circ}$.
    This PA misalignment is lower than the  accuracy with which it has been measured in \cygx and \scox,
     but it could represent a powerful scientific case for future higher sensitivity facilities.
\end{itemize}

The upper limit we put on the PD of the BL at high optical depths additionally shows that whenever model-dependent
spectropolarimetry attributes PD  values on the order of a few percent or more to the Comptonized component 
(see Sect. \ref{sect_introduction}), the exceeding part of the polarization
signal must be attributed to an extra-component, which provides a strong contribution in terms of polarimetry,
but a less significant one to the observed flux. Indeed, if the energy coverage is narrow, or the data statistics is not of high quality, a simple two-component continuum model is often enough to describe the observed spectrasatisfactorily, with the
contribution of other physical processes somewhat embedded by the Comptonization component.
The most natural candidate for the excess in the net polarized signal is reflection from the inner accretion disk 
illuminated by the radiation of the BL \citep[LS85;][]{coughenour2018, iaria2020, garcia2022}, with the possibility of achieving PD values as high as
50\% \citep{matt1993}; albeit, contribution from an outflowing subrelativistic
wind cannot be excluded \citep{dimarco2023, poutanen2023}. On the other hand, if disk reflection is detected but spectropolarimetry does not allow us to split contribution to total polarization over the single components unambiguously, our theoretical results provide strong
theoretical support to put tight constraints on at least one of them (the BL), helping to reduce
model degeneracy.

It is also important to point out that the results presented here implicitly assume albedo $A=0$ at the bottom of the slab
(see Fig.\ref{slab_geometry}).
Physically, this resembles the case of a neutral or poorly ionized material absorbing the back-scattered radiation of a surrounding ionized
atmosphere in an accretion disk, with the effect being dominant for photon energies $\la 10$ keV  \citep{mmc83, haardt91}.
For an NS photosphere illuminating from the bottom of the BL, at the characteristic temperature of the BB-like seed photon's
($\sim$ 1--2 keV) matter is ionized and partially reflective; thus, polarimetric and angular properties of reflected radiation need to be
 computed  either numerically \citep[C60;][]{basko74, mz95, ps96} or via Monte Carlo methods \citep{wlz88}.
In Fig. \ref{fig_ftop_vs_tau}, we report the fraction of flux at the top of the layer as a function of the optical depth 
of the slab. The behavior can be described by the functional form 

\begin{equation}
\varepsilon(2\tau_0)= e^{-a (2\tau_0)^b},
\end{equation}

\noindent
with $a= 0.72\pm0.013$ and $b=0.54\pm0.014$.
For progressively higher values of $2\tau_0$, an increasing fraction $1-\varepsilon(2\tau_0)$ of radiation is back-scattered to the bottom of the
layer, where it can be partially reflected.

Mathematically, the solution of Eqs. (\ref{initial_rte_a}) and (\ref{initial_rte_b}) for radiation reflected at the inner boundary can be obtained by substituting the source functions $S_{\rm l} (\tau, \mu)$ and $S_{\rm r} (\tau, \mu)$ 
with some terms that are functionals of the incident radiation, which we may express as 
$ \delta (\tau) I^{\rm refl}_{\rm l, r} (\mu) = \mathcal{F} [\Iminuslr (2\tau_0, \mu)]$.
The presence of a partially reflecting surface at the bottom of a scattering medium increases the average number of scattering
of photons before escaping, with a hardening of the Comptonization spectral slope \citep{tmk97}.
Polarization properties of emerging radiation can be affected as well.
However, for high optical depths, where diffusion approximation holds, photons lose information about their
initial angular distribution and polarization state (ST80, ST85). This is particularly true for photons initially located at the base of the
scattering atmosphere, such as Case 1 treated in the present work.
We performed a numerical check by using the crude approximation of seed photons initially polarized at a constant 20\% level
(see Figs. 24-26 of C60 for a more precise solution). 
For optical depths $2\tau_0 \ga 4$, the PD angular distribution resulted to have a  behavior qualitatively comparable to that shown in Fig. \ref{fig_pdegintensity_seed3}, with deviations in terms of PD value on the order of 15\% for $2\tau_0=4$, progressively decreasing for increasing $2\tau_0$. 
The deviations when passing to the observer frame integrating over the BL area of course resulted to be on the same order, in turn preserving the
 upper limit of 1\% obtained for unpolarized seed photons
 Thus, for albedo $A=0,$ results are valid for any value of the optical depth of the slab and the polarimetric properties of
emerging radiation are fully described by the solution of Eqs. (\ref{initial_rte_a}) and (\ref{initial_rte_b}).

On the other hand, when the albedo $A >0, $  two things can be deduced. Firstly For high optical depths ($2\tau_0 \ga 4$), the photon diffusion
regime is applicable; the total flux emerging from the slab is higher
    than the case $A=0$, but polarization properties and angular distribution
of emerging radiation do not vary significantly, independently of the polarization
status of the initial source function. 
    Secondly, as this is the typical optical depth of the Comptonization spectra of
 NS-LMXBs in a soft state, we conclude
    that results for the case shown in Fig. \ref{fig_bl_case1_tau5} are reliable. For low optical depths (less than a few), polarimetric properties
of radiation emerging at the top of the slab after reflection depends on
the initial state of the source function (i.e., reflected intensity at $\tau=0$),
which then must be computed
    carefully.
 
\begin{itemize}
    \item[] \end{itemize}

\section{Conclusions}\label{sect_conclusion}

In this paper, we presented results concerning the PD and PA of radiation scattered in a BL around an NS.
These quantities were obtained by first solving the coupled system of RTEs for the two polarization modes 
in the comoving frame for an electron-scattering atmosphere with plane-parallel geometry and Thomson  scattering cross-section, and then passing to the observer frame
with back-tracing light propagation in the \scw metric.
For seed photons located at the base of the slab, and typical optical depths of the Comptonization spectra of NS-LMXBs in a soft state, the PD is less than 0.5\%, while the PA is in the same direction as the NS spin axis, but it is slightly rotated by a few degrees;
the amount depends on the latitudinal extent of the BL as well as its rotational frequency.
For lower optical depths or seed photons uniformly distributed over the slab, the PA is instead at about a right angle with the same
amount of rotation.

The significantly higher values of the PD for the Comptonized component (BL) obtained with IXPE spectropolarimetric analysys of NS-LMXBs in the soft state so far
 are indicative that another process, presumably inner disk reflection, is responsible for it.
To our knowledge, this is the first time that an upper limit on the BL polarization is provided with a thorough numerical approach,
and the results presented here provide strong theoretical support for the interpretation of IXPE observations of NS-LMXB in the soft state,
where the contribution of the single components to the total polarization signal cannot be completely split when the model-dependent spectropolarimetric analysis is performed.

\begin{acknowledgements}
This work is carried out as part of the research funded by INAF Grant 2023 titled "Spin and Geometry in accreting X-ray binaries: The first multi frequency spectro-polarimetric campaign".
\end{acknowledgements}

\bibliographystyle{aa} 
\bibliography{biblio} 

\end{document}